\documentclass{emulateapj}
\usepackage{url,graphicx,amssymb}
\usepackage{subfigure,hyperref,lscape}

\long\def\symbolfootnote[#1]#2{\begingroup%
\def\thefootnote{\fnsymbol{footnote}}\footnote[#1]{#2}\endgroup}

\newcommand\qbh{GS 1354-64}

\begin{document}

\shorttitle{\qbh: An anomalous quiescent black hole}
\shortauthors{Reynolds et al.}

\title
{An Anomalous Quiescent Stellar Mass Black Hole}
\author{Mark T. Reynolds\altaffilmark{1}, Jon M. Miller\altaffilmark{1}}  
\email{markrey@umich.edu}

\altaffiltext{1}{Department of Astronomy, University of Michigan, 500 Church
  Street, Ann Arbor, MI 48109} 


\begin{abstract}
We present the results of a 40 ks \textit{Chandra} observation of the quiescent
stellar mass black hole \qbh. A total of 266 net counts are detected at the position
of this system. The resulting spectrum is found to be consistent with the spectra of
previously observed quiescent black holes, i.e., a power-law with a photon index of
$\rm \Gamma \sim 2$. The inferred luminosity in the 0.5 -- 10 keV band is found to lie
in the range $\rm 0.5 - 6.5 \times 10^{34}~erg~s^{-1}$, where the uncertainty in the
distance is the dominant source of this large luminosity range. Nonetheless, this
luminosity is over an order of magnitude greater than that expected from the known
distribution of quiescent stellar mass black hole luminosities and makes \qbh~the only
known stellar mass black hole to disagree with this relation. This observation
suggests the possibility of significant accretion persisting in the quiescent state.
\end{abstract}
 

\maketitle
\section{Introduction}
Observations at optical and infrared wavelengths have allowed us to measure the mass
of 19 Galactic and 4 extra-galactic stellar mass black holes ($\rm M_{BH} \sim 10 - 20
M\sun$:
\citealt{mcclintockremillard06,orosz07,silverman08,crowther10,corralsantana11}). The
presence of a black hole is inferred whenever the mass of the compact object is
dynamically constrained to be greater than $\rm 3~M_{\odot}$
\citep{rr74,kalogerabaym96}. This does \textit{not} confirm the compact object to be a
classical black hole and these observations could be interpreted to agree with more
exotic physics (see \citealt{narayan08} for details).  In order to confirm that the
objects above are indeed black holes, one is required to detect a unique
characteristic of such objects, i.e. the event horizon.

X-ray observations of accreting black holes have revealed a number of distinct
accretion regimes, which correlate in a broad sense with luminosity of the
source. The very-high, soft and hard states (hereafter VHS, SS \& HS) are the
  primary active accretion states observed in XRBs. At luminosities below that
observed in the HS state the system is said to be in the quiescent state.  The
quiescent state is characterized as an extremely faint state ($\rm L_x \leq
10^{33.5}~erg~s^{-1}$), with an X-ray spectrum that is distinctly non-thermal
($\Gamma$ = 1.5 -- 2.2; \citealt{mcclintockremillard06,corbel06}).

The quiescent X-ray emission from black hole binaries (BHBs) is not consistent with
expectations from standard accretion disc theory (e.g. \citealt{mcclintock95}).  The
current paradigm for understanding the X-ray emission from quiescent BHBs involves a
standard thin disc which transforms to a quasi-spherical inner flow at a distance of
$\rm \sim 10^3 –- 10^4$ Schwarzschild radii from the black hole. The inner flow
could consist of an advection dominated accretion flow (ADAF: see
\citealt{narayan08} and references therein).  In an ADAF the energy, released via
viscous dissipation, remains in the accreting gas rather than being radiated away. As
a result, most of the energy is advected with the accretion flow resulting in only a
small percentage of the energy being radiated by the gas before it reaches the compact
object. The radiative efficiency of such a flow is expected to be in the range $\sim$
0.01 -- 1\%. In comparison a standard thin-disc is expected to have an efficiency an
order of magnitude greater. The ADAF model has been successfully applied to
observations of a number of quiescent and hard state BHBs,
e.g. \citet{narayan96}. The ADAF solution is not limited to stellar
mass black holes and has also been successfully used to model the emission from a
number of supermassive black holes (SMBH), e.g. see \citet{yuan03} for an ADAF fit to
the spectrum of the least luminous known black hole Sgr A$^*$ ($\rm L_x \sim
10^{-9}~L_{Edd}$).

However, the ADAF solution also allows outflows as emphasized by \citet{bb99}. In
recent years jets/outflows have been recognized as a ubiquitous feature associated
with the process of accretion on the largest and smallest scales, e.g. active galactic
nuclei, XRBs.  \citet{fender03} proposed an alternative scheme, whereby at low
luminosity ($\rm L_x \lesssim 10^{-4}~L_{Edd}$) BHs should enter a 'jet-dominated'
state, in which the majority of the accretion power drives a radiatively-inefficient
jet. The detection of the black hole A0620-00 at radio wavelengths may support this
\citep{gallo06}, although the absence of a discernible jet in Sgr A$^*$ is problematic
\citep{narayan08}.  However, the detection of frequency dependent time-lags in the
radio flares from Sgr A$^*$ \citep{yusef09}, combined with the stratified
nature of the radio and mm-wave emission \citep{doeleman08} strongly suggest an
unbound, mildly relativistic outflow. Recent work envisages the X-ray flux from a
quiescent system to be a combination of an outflow/jet and an inner advective region,
e.g. \citet{yuan09}.

With the advent of the \textit{Chandra} \& \textit{XMM-Newton} X-ray observatories
detailed observations of quiescent black hole and neutron star binary systems have
become possible, i.e. $\rm L_x \lesssim 10^{-6}~L_{Edd}$. As first pointed out in
\citet{narayan97} \& \citet{garcia98}, the observed luminosities of the
quiescent BHs are systematically fainter than NSs. For a black hole the energy stored
in the flow is advected across the event horizon whereas for a neutron star the
material strikes the solid surface where it is re-radiated \citep{garcia01,kong02}.
Subsequent observations have confirmed this picture
\citep{sutaria02,hameury03,tomsick03,mcclintock04,corbel06,gallo08}.

\qbh~was discovered by \textit{Ginga} in 1987, where it displayed an X-ray spectrum
dominated by a soft blackbody disc component, consistent with that of the then known
black hole binaries in the soft state \citep{makino87,kitamoto90}. A second outburst
was observed from this system in 1997 by \textit{RXTE}, though on this occasion the
system remained in the low-hard state throughout the outburst \citep{brocksopp01}. The
black hole nature of the primary has been dynamically confirmed by
\citet{casares04,casares09}, who found the system to comprise of a G0-5 III mass donor
in a $\sim$ 2.5 days orbit around the black hole. The measured radial velocity ($\sim$
279 km s$^{-1}$) in combination with the absence of any X-ray eclipses sets a secure
lower limit for the mass of the black hole of $\rm M_x \geq 7.6 \pm
0.7~M_{\sun}$. \qbh~is found to lie at a large distance, where 25 kpc $\leq$ d $\leq$
61 kpc, and the upper limit is obtained by requiring the 1987 outburst to be
sub-Eddington. This makes \qbh~ the most distant known Galactic black hole.  The field
containing \qbh~has not been previously observed by either \textit{XMM-Newton} or
\textit{Chandra}. The only X-ray imaging of this field was carried out by
\textit{Rosat} as part of the all sky survey \citep{voges99}, where only a weak upper
limit exists, $\rm f_x \leq 5.6 \times 10^{-13}~erg~s^{-1}~cm^{-2}$.

In this letter, we describe an observation of the quiescent Galactic stellar mass
black hole binary \qbh~with the \textit{Chandra X-ray Observatory}. We describe the
observations and extraction of source spectra and lightcurves, these data are
discussed in the context of models for the quiescent accretion flow.

\begin{figure}[t]
\begin{center}
\includegraphics[height=0.34\textheight,angle=-90]{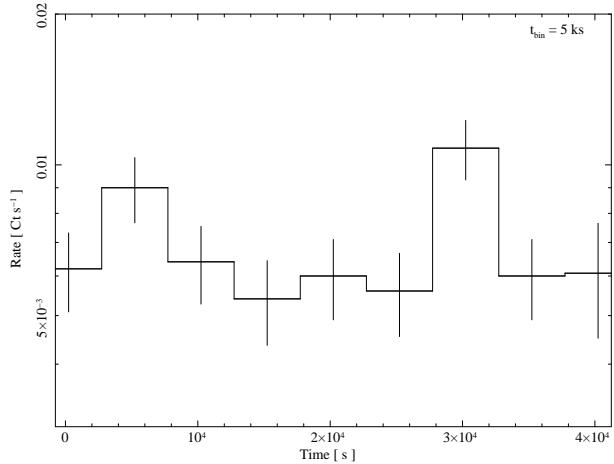}
\caption{\qbh~background subtracted lightcurve as observed by \textit{Chandra}, where
  the lightcurve has been divided into 5 ks bins. Variability of a factor of two is
  observed on kilo-second timescales.}
\label{5ks_lc}
\end{center}
\end{figure}

\section{Observations}
\qbh~was observed by \textit{Chandra} for 40 ks on 2010 October 1 (MJD 55470,
obsid: 12471, PI: Reynolds), where it was placed on the back illuminated ACIS-S3
detector, which was operated in \textsc{vfaint} mode. A significant point source is
detected consistent with the known position of \qbh~\citep{brocksopp01}.  As the I3,
S1--S4 detectors were active during this observation, a number of serendipitous point
sources are also detected. These sources (31 in total) and their properties will be
described in a separate publication.

The \qbh~spectrum was extracted from a 2.5\arcsec~region with \texttt{psextract},
which also generated the appropriate response files. A background spectrum was
extracted from a neighbouring source free region on the detector. We detect 266 net
counts consistent with the known position of \qbh. The source \& background
lightcurves were extracted using \texttt{dmextract} and binned using
\texttt{lcurve}. Inspection of the background lightcurve reveals an absence of
  any significant flares during the observation. The background subtracted lightcurve
is displayed in Fig. \ref{5ks_lc}, where it has been divided into 5 ks
bins. Variability of a factor of 2x is observed on kilo-second timescales, e.g., the
count rate is observed to increase from an average of $\rm \sim 0.006~ct~s^{-1}$ to
$\rm \sim 0.012~ct~s^{-1}$ at the 30 ks mark.

All data reduction and analysis takes place within the \textsc{heasoft 6.6.2}
environment, which includes \textsc{ftools 6.6, ciao 4.2} and \textsc{xspec
  12.5.0aj}. The latest versions of the relevant \textit{Chandra} \textsc{caldb} files
are also used.

\section{Analysis}
\begin{figure}[t]
\begin{center}
\includegraphics[height=0.34\textheight,angle=-90]{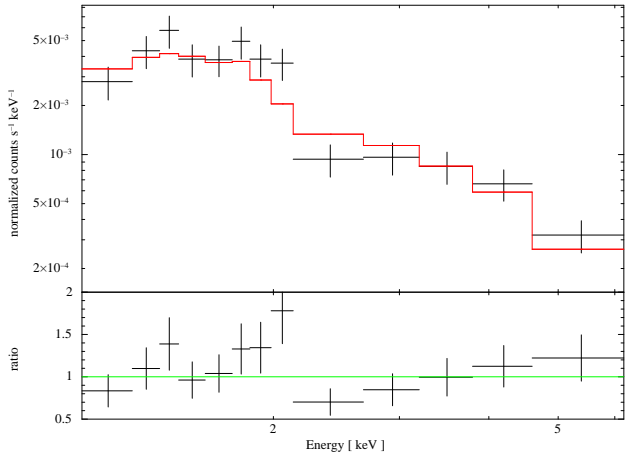}
\caption{\textit{Chandra} 1 -- 6 keV spectrum of \qbh. The exposure time is $\sim$ 39
  ks. The best fit power-law model ($\Gamma \sim 2.1$) is plotted, the source
  luminosity is $\rm L_x \sim 0.7-6 \times 10^{34}~erg~s^{-1}$, see Table
  \ref{spec_params} for details.}
\label{spec_fit}
\end{center}
\end{figure}

The relatively large number of counts detected facilitates basic spectral fitting. The
spectrum was binned such that each bin contained 20 counts using \texttt{grppha},
providing useful data in the spectral range 1 -- 6 keV. We model the spectrum using 2
models for the continuum (i) a power-law (\texttt{pha*po}), and (ii) Bremsstrahlung
(\texttt{pha*brem}). The results of the model fits are displayed in
Table. \ref{spec_params}. The best fit power-law spectral index of $\rm \Gamma
  \sim 2.1$ (see Fig \ref{spec_fit}) and the bremsstrahlung temperature of $\rm T_B
  \sim 5~keV$ are consistent with previous observations of quiescent stellar mass black
  holes, e.g., \citet{kong02,bradley07}. We measure an absorbed flux of $\rm \sim
7.7(7.4)\times 10^{-14}~erg~s^{-1}~cm^{-2}$ in the 0.5 -- 10.0 keV band for the power-law(bremsstrahlung) model. This corresponds
to an unabsorbed flux of approximately $\rm \sim 1.5(1.1)\times
10^{-13}~erg~s^{-1}~cm^{-2}$ in the same energy range, see Table \ref{spec_params} for
details.

\begin{figure}[t]
\begin{center}
\includegraphics[height=0.34\textheight,angle=-90]{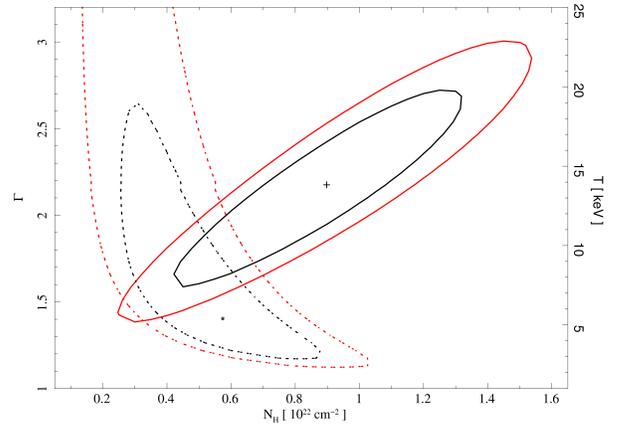}
\caption{Contour plots corresponding to the best fit models in Table
  \ref{spec_params}. The power-law fit is represented by the solid lines, while the
  thermal bremsstrahlung model is represented by the dotted contours. Contours plotted
  represent the 68\% and 90\% confidence levels.}
\label{contour_plots}
\end{center}
\end{figure}

In Fig. \ref{contour_plots}, we plot contours of power-law index and
bremsstrahlung temperature versus extinction for each of the models above. The
uncertainty regions are large consistent with the small energy range covered by the
data. Nonetheless, we see that the power-law model favours significantly larger values
for the extinction $\rm N_H \sim 9\times 10^{21}~cm^{-2}$. For comparison,
\citet{kitamoto90} estimated E(B-V) $\rm \sim 1~(N_H \sim 5.3\times 10^{21}~cm^{-2}$),
and the extinction estimated from the optical spectra obtained by
\citet{casares04,casares09} are consistent with this value.

As the measured spectrum is intrinsically soft, the differing extinction estimates
result in the unabsorbed flux for the power-law model being approximately 1.5 times
that measured using the bremsstrahlung model. The power-law model is formally a better
fit to the observed spectrum; however, the low number of bins render this
insignificant.  

In Fig. \ref{qbh_luminosity}, we plot the luminosity calculated using the unabsorbed
flux from the best fit power-law model in Table \ref{spec_params} and the distance
estimates of \citet{casares09}, i.e., 25 $\rm \leq d_{kpc} \leq$ 61. \qbh~is observed
to be at a luminosity over an order of magnitude greater than that expected from a
quiescent black hole. For this system to be consistent with the known distribution of
black holes would require a decrease of the distance to $\sim$ 5.8 kpc. Monitoring
observations at optical wavelengths with the Faulkes
telescope\footnote{\url{http://staff.science.uva.nl/~davidr/faulkes/}\citep{lewis08}}
revealed \qbh~to be consistent with the known quiescent optical flux at the time of
our \textit{Chandra} observations, though, we note the nearest observation took place
approximately 2 months prior to the \textit{Chandra} observation described
herein. While observations with the available X-ray all sky monitors show no evidence
for increased activity either before or after our observation. However, a faint
outburst would not have been detected by these observations.

There are 2 other stellar mass black holes with weak quiescent luminosity constraints.
The first system, H 1705-250, had only been observed by \textit{Rosat}. This upper
limit will be updated soon by \textit{Chandra} (Obsid: 11041, PI: Kong). The second
system, GX 339-4, is an extremely active transient having exhibited 5 large and $\sim$
10 small outburst in the last 15 years. The most sensitive observation was obtained by
\textit{Chandra} in 2003 and revealed this system at the faintest level measured to
date. However, this observation occurred $\sim$ 10 months after the 2002 outburst and
7 months prior to the 2003/2004 outburst \citep{gallo03}. Hence, it is highly unlikely
that GX 339-4 was in quiescence at the time of this observation. The current upper
limits for these systems are represented by the solid triangles in
Fig. \ref{qbh_luminosity}.

\section{Discussion}
We have observed the quiescent stellar mass black hole \qbh~with \textit{Chandra} and
found it to be at a luminosity of $\rm \sim 10^{34}~erg~s^{-1}$. This is over an order
of magnitude greater than expected. Here, we discuss the uncertainty in the luminosity
estimate and consider the implications of this measurement for the nature of the
accretion flow in the quiescent state. 

\subsection{Distance to \qbh}\label{distance}
In order to compare to the other systems in Fig. \ref{qbh_luminosity}, we must know
the distance to each system. \citet{casares04,casares09} have calculated the minimum
distance to \qbh~to be $\rm d_{min} \geq 25~kpc$. The distance estimate depends on the
known spectral type of the secondary star, the apparent magnitude of the star, the
contribution from the accretion disk and the extinction. The extinction is the most
uncertain quantity in the distance calculation. \citet{casares04} estimate
the extinction from known empirical relationships between observed spectral features and
the column density resulting in an estimated E(B-V) $\sim$ 1, or more specifically
E(B-V) $\geq$ 0.78, consistent with the estimate of \citet{kitamoto90}.   
However, the measured neutral hydrogen column in this field suggests a larger value,
i.e., $\rm N_H \sim 7.27\times 10^{21}~cm^{-2}$ \citep{kalberla05} or E(B-V) $\sim$
1.4. 

A higher level of extinction will effect the estimated minimum distance to this system
of $\rm d_{min} \geq 25~kpc$, in particular for a column density of $\rm N_H \sim
7.27\times 10^{21}~cm^{-2}$ we find a minimum distance of $\rm d_{min} \geq 17.8~kpc$,
decreasing to 10.2 kpc for a column of $\rm 10^{22}~cm^{-2}$.  The empirical diffuse
interstellar band relation of \citet{herbig75} as used in \citet{casares04} suggests
an upper limit to the reddening of E(B-V) $\leq$ 1.5 ($\rm N_H \sim 7.9\times
10^{21}~cm^{-2}$) corresponding to a distance of $\sim$ 15.6 kpc.  Nonetheless, even
in the unlikely event that the extinction is underestimated, \qbh~ remains too
luminous in comparison to the known distribution.

\subsection{Possibilities}
The measured X-ray luminosity allows one to estimate the mass accretion rate, we do
this under 2 assumptions for the accretion efficiency, i.e., $\rm L_x = \eta
\dot{M}c^2$. Assuming standard thin disc accretion ($\rm \eta = 0.1$) implies an
accretion rate of $\rm 10^{-11}~(L_x/10^{34})~M_{\sun}~yr^{-1}$, whereas for an ADAF
an efficiency of $\rm \eta = 10^{-2} - 10^{-4}$ is expected, and hence, a
commensurably larger accretion rate. Accretion rates onto the outer disc of $\sim \rm
10^{-10}~M_{\sun}~yr^{-1}$ have been estimated for a number of quiescent black hole
binaries, e.g., A0620-00 \citep{mcclintock95} and XTE J1550-564 \citep{orosz11}.  As
such, the accretion rates necessary to generate the observed luminosity appear to be
feasible. 

\begin{figure}[t]
\begin{center}
\includegraphics[height=0.28\textheight]{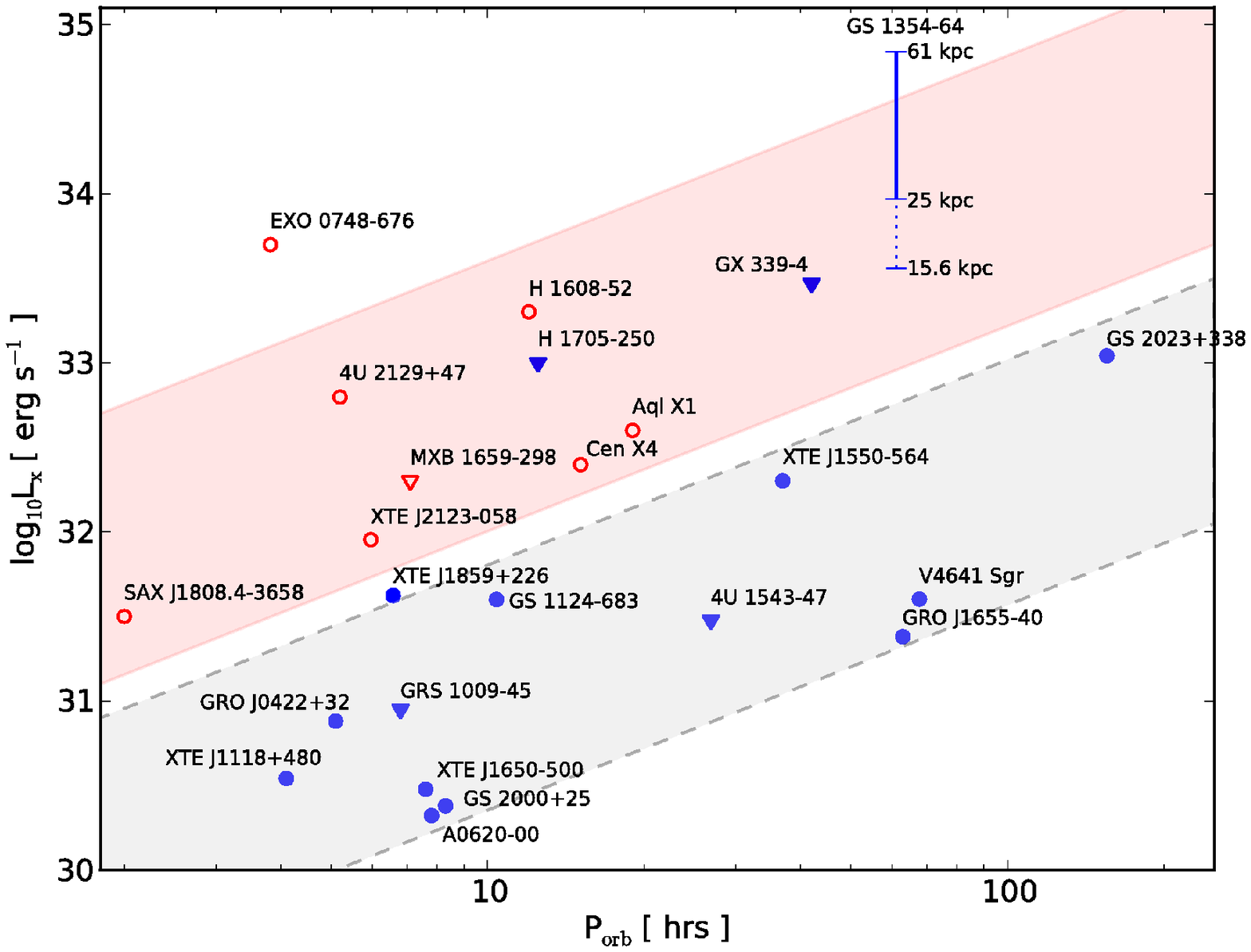}
\caption{The vertical blue line indicates the measured luminosity of \qbh~
    assuming a distance of $\rm 25~kpc \leq d \leq 61~kpc$ \citep{casares09}, while
    the dotted extension indicates the estimated lower limit to the distance, see \S
    \ref{distance} for details. \qbh~is observed to be over an order of magnitude too
    luminous in comparison to the known distribution of quiescent black holes (solid
    blue symbols). The open red symbols indicate the quiescent luminosities of neutron
    stars. Upper limits are represented by triangles. The red and blue filled areas
    indicate the regions in the $\rm L_x,~P_{orb}$ plane occupied by the neutron star
    and black holes systems respectively. Black hole luminosities are from
    \citet{garcia01,kong02,sutaria02,tomsick03,hameury03,mcclintock04,corbel06,gallo08},
    and the neutron star luminosities are from
    \citet{garcia01,tomsick04,lin09,degenaar11}.}
\label{qbh_luminosity}
\end{center}
\end{figure}

We also note the similarity between the observed X-ray variability of a factor of
$\sim$ 2 (see Fig. \ref{5ks_lc}) and that typically observed at optical wavelengths,
where variability of approximately 1 magnitude ($\sim$ 2.5x in linear units) is common
from \qbh~in quiescence \citep{casares04,casares09}. Correlated optical/X-ray
variability has been observed from V404 Cyg in quiescence. However, the origin of this
variability is not understood, with emission from the inner ADAF and/or irradiation of
the disk being likely \citep{hynes04}. A similar scenario is likely here, in the case
of \qbh; however, further observations are required to confirm this possibility.

Below we briefly discuss what we consider to be the most likely explanation
for the large observed quiescent luminosity:

\noindent \textit{(i) A temporary increase in the mass accretion rate, $\rm \dot{M}$.}
For example, observations of the stellar mass black hole GRO J1655-40 ($\rm P_{orb}
\sim 63~hrs$) have revealed luminosity variations by a factor of 10
\citep{hameury03}. However, the brighter of these observations was obtained $\sim$ 9
months after the 1995 outburst and only one month prior to the 1996 outburst, i.e.,
the observed large luminosity was caused by the system not being quiescent at the time
of the \textit{ASCA} observations. Long term variations of a factor of $\sim$ 10 have
also been observed from V404 Cyg ($\rm P_{orb} \sim 155~hrs$, \citealt{bradley07}),
while variations of $\sim$ 20x have been observed on kilo-second timescales with
\textit{Chandra} \citep{hynes04}. However, in both cases the luminosity remained
consistent with that expected for the distribution of quiescent black holes, see
Fig. \ref{qbh_luminosity}.  Observations of Sgr A$^*$ have revealed flares with an
amplitude of greater than 100 times the true quiescent rate
\citep{baganoff01,porquet08}, but these flares typically have a duration of a few
hours and as such are unlikely the cause of the large luminosity, which we observe
from \qbh.

\noindent \textit{(ii) The actual quiescent accretion luminosity from this system is
  $\rm \gtrsim 10^{34}~erg~s^{-1}$.}  This would be the largest Eddington scaled
luminosity measured from a black hole in quiescence to date, and would point to the
existence of a unique low luminosity accretion flow in this system. Standard accretion
disc theory predicts the disk to be truncated, with a low mass accretion rate and
luminosity \citep{mcclintock95,narayan96,lasota08}, e.g., both GRO J1655-40 ($\rm M_x
\sim 6.5~M_{\sun},~P_{orb} \sim 63~hrs$) \& V4641 Sgr ($\rm M_x \sim 7~M_{\sun}, P_{orb}
\sim 68~hrs$) have similar orbital periods to \qbh~($\rm M_x \gtrsim 7~M_{\sun}, P_{orb}
\sim 60~hrs$), but have quiescent luminosities consistent with the other black holes,
i.e., $\rm L_x \sim 5.9\times 10^{31}~erg~s^{-1}$, $\rm 4\times 10^{31}~erg~s^{-1}$
respectively \citep{hameury03,tomsick03}. The nature of the mass donor secondary star
is also sufficiently similar with $\sim$ F4III-IV \citep{orosz97} and $\sim$ B9III
\citep{orosz01} respectively in comparison to the G0-G5III secondary in \qbh. If the
large quiescent luminosity we have measured is confirmed, it would suggest the
existence of a previously unrecognized stable mode of low luminosity accretion ($\rm
L_x \sim 10^{-5}~L_{Edd}$), and with it a population of relatively faint accreting
black holes in the galaxy, e.g., \citet{menou99}.

\subsection{BH vs NS luminosities}
A comparison of the quiescent luminosities of the black hole and neutron star binaries
revealed the black hole systems to be a factor of $\sim$ 100 times fainter
(\citealt{garcia01}, Fig. \ref{qbh_luminosity}). This has been interpreted as evidence
for the absence of a solid surface in the black hole systems, and conversely as
indirect evidence for the existence of the event horizon. The observation presented of
\qbh~herein contradicts this empirical relationship. Previously, an observation of the
neutron star binary 1H 1905+00 was claimed to also contradict this relationship
\citep{jonker07}; however, further analysis revealed this to be false
\citep{lasota08}. If the measured luminosity of \qbh~(Fig. \ref{qbh_luminosity}) is
shown to be stable by future observations, one could ask if the observed difference
between the black hole and neutron stars in quiescent is not caused by the absence of
a surface in the black hole systems but instead by the favoring of this higher
luminosity quiescent accretion flow in the neutron star systems.

\begin{table*}
\begin{center}
\caption{Broadband continuum fit parameters}\label{spec_params}
\begin{tabular}{lccccccc}
\tableline\\ [-2.0ex]
Model & $\rm N_H$ & $\Gamma$ & kT & $\rm f_{x~abs} (0.5-10.0~keV)$ & $\rm f_x
(0.5-10.0~keV)$ & $\rm L_{x} (25 kpc, 61 kpc)$ & $\rm \chi^2/\nu$\\ [0.5ex]  
& $\rm [~10^{22}~cm^{-2}~]$ & & [ keV ] & $\rm [~10^{-14}~erg~s^{-1}~cm{-2}~]$ & $\rm
[~10^{-14}~erg~s^{-1}~cm{-2}~]$  & $\rm [~10^{33}~erg~s^{-1}~]$ & \\ [0.5ex]
\tableline\tableline\\ [-2.0ex] 
\texttt{pha(po)}     & 0.9$^{+0.45}_{-0.4}$ & 2.2$^{+0.6}_{-0.4}$ & -- &
7.69$^{+0.9}_{-0.7}$ & 14.5$^{+1.7}_{-1.3}$ & 7.65$^{+0.90}_{-0.68}$,
64.6$^{+7.5}_{-5.8}$  & 14.3/10 \\   [0.5ex]
\texttt{pha(bremss)}     & 0.58$^{+0.34}_{-0.28}$ & -- & 5.1$^{+7.3}_{-2.4}$ &
7.4$^{+0.9}_{-0.7}$  & 10.7$^{+1.0}_{-1.2}$ & 5.65$^{+0.52}_{-0.64}$,
47.6$^{+4.5}_{-5.3}$ & 15.7/10 \\   [0.5ex]  
\tableline
\end{tabular}
\tablecomments{Best fit model parameters for the \qbh~continuum as measured on the
  ACIS-S3 detector in the spectral range 1 -- 6 keV, see Fig. \ref{spec_fit}. The flux
  is calculated in the 0.5 -- 10.0 keV band, and the luminosity has been calculated
  assuming a distance of $\rm 25 \leq d_{kpc} \leq 61$ \citep{casares09}. All errors
  are quoted at the 90\% confidence level.}
\end{center}
\vspace{2mm}
\end{table*}

\medskip

\acknowledgements 
We thank Dave Russell for discussions regarding the Faulkes monitoring of this system.
MTR \& JMM gratefully acknowledge support through the \textit{Chandra} Guest Observer
program. This research made use of the \textit{SIMBAD} database, operated at CDS,
Strasbourg, France and NASA's Astrophysics Data System.  We thank the anonymous
referee for useful comments


\vspace{1cm}
\footnotesize{This paper was typeset using a \LaTeX\ file prepared by the 
author}


\end{document}